\documentclass[twoside]{article}
\usepackage{qic,graphics, chapterbib}

\renewcommand{\thefootnote}{\fnsymbol{footnote}}  

\begin{document}
\setlength{\textheight}{8.0truein}    

\runninghead{Scalable, efficient ion-photon coupling with phase Fresnel lenses for large-scale quantum computing}
            {Authors E. W. Streed et al }

\normalsize\textlineskip
\thispagestyle{empty}
\setcounter{page}{1}


\vspace*{0.88truein}

\alphfootnote

\fpage{1}

\centerline{\bf
Scalable, efficient ion-photon coupling with phase Fresnel lenses for large-scale quantum computing}
\vspace*{0.035truein}

\centerline{\footnotesize
E.W. Streed, B.G. Norton, J.J. Chapman, and D. Kielpinski}
\vspace*{0.015truein}
\centerline{\footnotesize\it Centre for Quantum Dynamics, Griffith University}
\baselineskip=10pt
\centerline{\footnotesize\it Nathan, QLD 4111, Australia}
\vspace*{10pt}

\vspace*{0.21truein}
\abstracts{
Efficient ion-photon coupling is an important component for large-scale ion-trap quantum computing. We propose that arrays of phase Fresnel lenses (PFLs) are a favorable optical coupling technology to match with multi-zone ion traps. Both are scalable technologies based on conventional micro-fabrication techniques. The large numerical apertures (NAs) possible with PFLs can reduce the readout time for ion qubits. PFLs also provide good coherent ion-photon coupling by matching a large fraction of an ion's emission pattern to a single optical propagation mode (TEM$_{00}$). To this end we have optically characterized a large numerical aperture phase Fresnel lens (NA=0.64) designed for use at 369.5 nm, the principal fluorescence detection transition for Yb$^+$ ions. A diffraction-limited spot $w_0=350\pm15$ nm ($1/e^2$ waist) with mode quality $M^2= 1.08\pm0.05$ was measured with this PFL. From this we estimate the minimum expected free space coherent ion-photon coupling to be 0.64\%, which is twice the best previous experimental measurement using a conventional multi-element lens. We also evaluate two techniques for improving the entanglement fidelity between the ion state and photon polarization with large numerical aperture lenses.}{}{}
\vspace*{10pt}

\vspace*{3pt}

\vspace*{1pt}\textlineskip    
%
%
%
%

\setcounter{footnote}{0}
\renewcommand{\thefootnote}{\alph{footnote}}

\section{Introduction}
\begin{figure} [htbp]
\centerline{\includegraphics{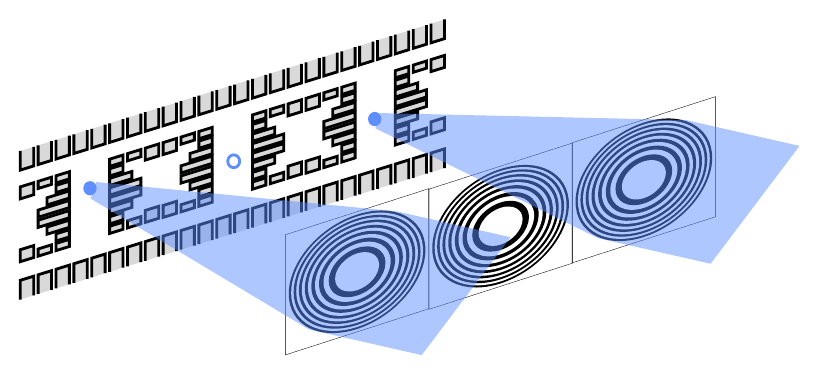}}
\vspace*{13pt}
\fcaption{\label{Fig1} Schematic of parallel optical operations on a multi-zone ion chip trap using an array of phase Frensel lenses.  }
\end{figure}

Quantum information processing leverages properties of quantum physics to perform computational and communications tasks at faster rates \cite{Shor-94, Grover-97} or with greater security \cite{Bennett-84} than classical techniques. Interest in this area has been stimulated by Shor's algorithm \cite{Shor-94} for efficient factoring of large numbers, since modern public key encryption schemes rely on the intractability of this problem with classical computational algorithms. The electronic and motional states of trapped ions are one of the leading systems for realizing quantum information processing. Trapped ions have long coherence times, strong yet controllable inter-qubit coupling, and are easy to prepare, manipulate, and read out using established optical and microwave techniques. Many small-scale quantum computation tasks have been demonstrated with trapped ions \cite{Kielpinski-03, Leibfried-07, Benhelm-08, Brickman-05, Blinov-04} and a roadmap exists for larger scale architectures \cite{Kielpinski-02, Kim-05, Steane-07}. A common thread in all the proposed large scale ion trap quantum computing architectures is the need for a scalable, efficient method for collecting ion fluorescence. Arrays of phase Fresnel lenses (PFLs) are well suited to meeting these requirements because of their large numerical apertures and scalable production via conventional micro-fabrication techniques. Fig. \ref{Fig1} illustrates the integration of a PFL array with a multi-zone ion trap to create a high-density quantum processor.

PFLs are diffractive optical elements manufactured using conventional electron beam nano-lithographic techniques. PFLs achieve diffraction-limited performance at high numerical apertures because on-axis geometrical aberrations are completely removed as part of the design process.  Diffraction-limited performance with large numerical apertures (NA=0.9, 28\% coverage of the total solid angle) has been demonstrated \cite{Menon-06} in the near UV. At large NAs ($>$  ~0.5) the ion emission pattern ceases to resemble a point source and exhibits polarization dependent structure. PFLs offer design flexibility for mode matching between a specific emission pattern and a single optical spatial mode (TEM$_{00}$), maximizing the coherent coupling. While the diffraction efficiency of a high-NA multilevel PFL was previously thought to be limited to ~20\% at deflection angles near 45$^\circ$, recent  vector diffraction modeling of PFLs  \cite{Cruz-Cabrera-07} shows efficiencies of 63\% could be obtained in this regime with a modified groove structure. The modeling also indicates that coating PFLs with a 20 nm layer of indium tin oxide to make them more electrically compatible with the ion trap environment would reduce the diffraction efficiency by only 12\%. In this paper we characterize the optical properties of an NA=0.64 phase Frensel lens. This PFL will be inserted into a Yb$^+$ ion trap system for proof of concept demonstration. From these measurements we calculate the expected coherent coupling efficiency between the spontaneous emission from single ion and a fundamental gaussian mode (TEM$_{00}$), which is equivalent to the efficiency for coupling into a cavity or a single mode fiber. We also evaluate potential limitations specific to PFLs for several atom-photon entanglement schemes. In addition, we propose two solutions to entanglement fidelity limitations arising from the use of high-NA collection optics in an especially useful atom-photon entanglement scheme.

%
%
%
%

\section{Experimental}
\noindent

\begin{figure} [htbp]
\centerline{\includegraphics{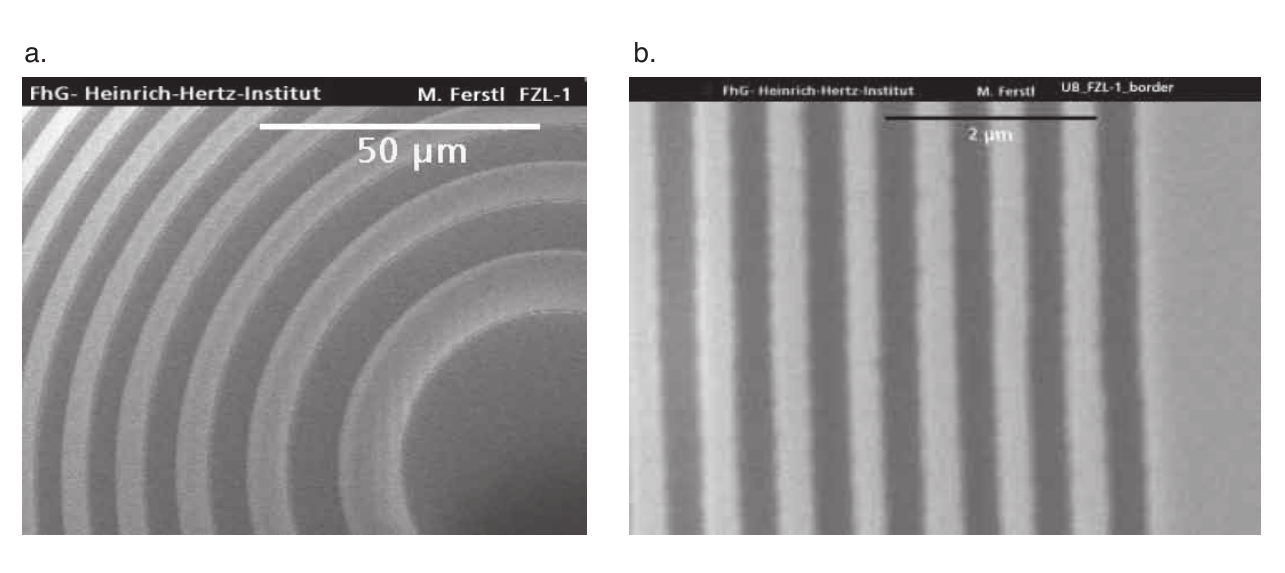}} 
\vspace*{13pt}
\fcaption{Electron microscopy images of the patterned phase Fresnel lens surface. a. Center region showing the innermost rings. b. Edge region with structures of comparable size to the $\lambda=369. 5$ nm design wavelength. Images courtesy of M. Ferstl, Heinrich-Hertz-Institut of the Fraunhofer-Institut f\"{u}r Nachrichtentechnik.}
\label{FigSEMPicsOfPFL}
\end{figure}

The PFL (Fig. \ref{FigSEMPicsOfPFL}) was fabricated by electron-beam lithography on a fused silica substrate at the Fraunhofer-Institut f\"{u}r Nachrichtentechnik in Germany. The e-beam patterning defined series of rings of radius $r_p^2= 2 f p \lambda + p^2 \lambda^2$, corresponding to contours with a $\pi$ phase step, according to the scalar design equation for a binary PFL. Here $f = 3$ mm is the design focal length,p is the ring index number,  and $\lambda$ = 369.5 nm, the design wavelength, is the wavelength of the S$_{1/2}$-P$_{1/2}$ cycling transition in Yb$^+$. The rings were etched to a depth of 390 nm, shifting the optical path length in the etched zones by half a wavelength. The lens clear aperture of 5 mm gives the lens a speed of F/0.6 and a NA=0.64, which corresponds to 12\% of the total solid angle since $NA\equiv \sin{\theta}$ and so $NA=\frac{1}{1+ 4(F/\#)^2}$. For small NA ( $\theta< ~0.3$ ) the approximation $NA \approx \frac{1}{2 F/\#}$ is often used, but is not valid in the high NA regime of interest. Aspheric lenses are often specified using the approximate NA formula even outside its range of validity, leading to grossly overstated catalog NA values.

\begin{figure} [htbp]
\centerline{\includegraphics{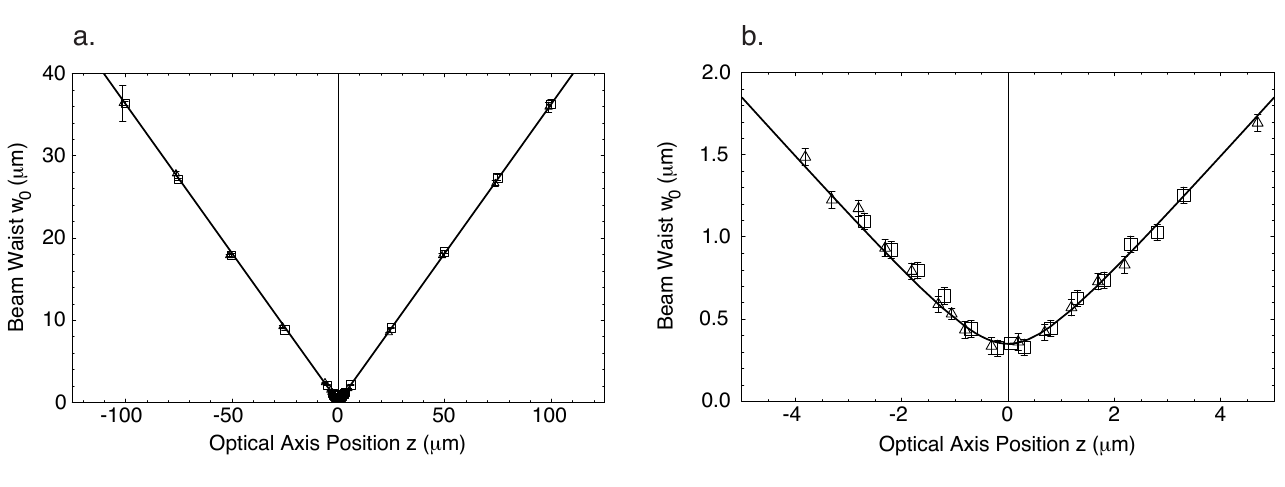}} 
\vspace*{13pt}
\fcaption{Focusing performance of a large aperture ( NA=0.64 ) binary phase Fresnel lens, 5 mm diameter clear aperture and 3 mm focal length with a 2.2 mm diameter input beam. a. Beam waist size as a function of position along the optical axis (z).  b. Detail of focusing region from a. Fit curve is to Eq. \ref{eqBeamWaist} with $w_0=350\pm15$ nm and $M^2= 1.08\pm0.05$. The half angle of the beam divergence is $\theta=348\pm1$ mrad and the nominal Rayleigh range $z_R=\pi w_0^2/\lambda=1040\pm90 nm$. Data was taken with the knife edge moving in (triangles) and out (squares) of the beam. An imperfection in the translation stage resulted in a systematic shift of $1.11\pm0.05$ $\mu$m between the z location of the in and out curves. This artifact has been removed from the plotted data. }
\label{FigBeamWaistData}
\end{figure}

Using a previously developed sub-micron beamprofiler \cite{Chapman-07}, the focusing properties of this NA=0.64 binary PFL were measured using the knife edge technique. The focused beam is chopped by a razor blade oriented perpendicular to the optical axis and the optical power transmission as a function of position is fitted to determine the beam size. We define the beam waist $w_0$ as the $1/e^2$ intensity radius. Sub-micron accuracy in measuring the beam waist is realized with monitoring of the razor blade position by a Michelson interferometer. Fig. \ref{FigBeamWaistData} shows a series of beam size measurements near the focus of the PFL given an input beam at the design wavelength and an input beam waist of 1.1 mm. The data was fitted to Eq. \ref{eqBeamWaist}  resulting in a minimum beam waist  of $w_0=350\pm15$ nm and a beam propagation factor $M^2=1.08\pm0.05$, indicating almost ideal gaussian behavior. The fitted beam waist value $w_0$ was also verified against the actual experimental data to prevent inadvertent "false fitting" of an unobserved lower beam waist. The beam propagation factor $M^2$ represents the increase in beam divergence over that for an ideal gaussian beam of equal waist size and is defined through the equation \cite{Siegman-98}

\begin{equation}
w^2(z)=w_0^2 + M^4 \times \left( \frac{\lambda}{\pi w_0} \right)^2 z^2
\label{eqBeamWaist}
\end{equation}

The overall diffraction efficiency of the PFL into the focus was measured to be $\bar{\eta}_{\mbox{diff}}=30\pm1$\% of input power, comparable to the ideal efficiency of 37\% (including Fresnel reflection losses) for a binary phase Fresnel lens. The total transmission was $92\pm1$\%, in agreement with the expected losses from Fresnel reflections (4\% per surface) for a flat fused silica plate.

%
%
%
%

\section{Analysis}
\subsection{Coupling efficiency}
\noindent
In an ion-trap quantum computer, light from an ion is either collected and detected or coupled into a subsequent optical device such as a single mode fiber, Fabry-Perot cavity, or interferometer. The probability that a photon is successfully collected ($p_{coll}$, Eq. \ref{eqCollection} ) by a lens depends on its average efficiency ($\bar{\eta}_{\mbox{diff}}$), numerical aperture (NA), the transition polarization ($\sigma$ or $\pi$), and the viewing orientation. The beam quality produced by the coupling optic is unimportant in photon counting applications so long as the collected light falls on the detector's active area. However, the probability that light from an ion is coherently coupled into a single optical mode ($p_{coh}$) does depend on the spatial quality of the beam (beam propagation factor $M^2$) which can be obtained for a particular beam divergence $\theta$. The coherent ion-photon coupling can be estimated by approximating the measured beam as an ideal gaussian, normalizing the intensity with the top hat approximation, and applying this effective divergence angle $\theta_e= \theta / ( M \sqrt{2} )$ to a polarization and orientation dependent formula for the fraction of light emitted into a cone. The actual beam can be approximated as an ideal gaussian with its divergence angle reduced by $1/M$ from the measured divergence $\theta=348\pm1$ mrad. The coherent coupling $p_{coh}$ of a spontaneously emitted photon into a single TEM$_{00}$ optical mode is calculated according to Eq. \ref{eqCoherCoupling}, where $\bar{\eta}_{\mbox{diff}}$ is the overall diffraction efficiency. The error introduced by applying the top hat approximation is less than 2\% for divergences of less than 0.93 radians ( NA$< 0.8$ for a beam diameter equal to the clear aperture). This is in contrast to the collection efficiency (Eq. \ref{eqCollection}) which depends only on the maximum collection angle (or NA) and the overall diffraction efficiency.

 \begin{equation}
p_{coll}= f_{\mbox{orient, pol}}\left(\theta_{\mbox{max}} \right) * \bar{\eta}_{\mbox{diff}}
\label{eqCollection}
\end{equation}

 \begin{equation}
p_{coh}= f_{\mbox{orient, pol}}\left(\theta \frac{1}{M} \frac{\sqrt{2}}{2}\right) * \bar{\eta}_{\mbox{diff}}
\label{eqCoherCoupling}
\end{equation}

\begin{figure} [htbp]
\centerline{\includegraphics{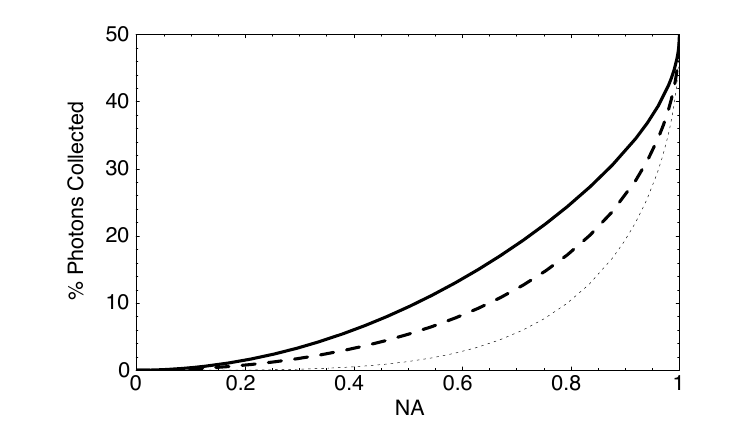}} 
\vspace*{13pt}
\fcaption{Fraction of light captured as a function of lens NA. Solid black line (upper) is for a $\pi$ transition oriented along equator (optical axis perpendicular to magnetic field) and $\sigma$ transitions oriented along the pole (optical axis parallel with magnetic field). Dashed line (middle) is $\sigma$ light from equatorial orientation, Dotted line (lower) is  $\pi$ transition light from polar orientation. }
\label{FigLensNACaptureFraction}
\end{figure}

We have calculated the emission collection fraction $f(\theta_m)$ as a function of acceptance angle for two different optical orientations (Fig. \ref{FigLensNACaptureFraction}); the magnetic field parallel to the optical axis (a polar view in spherical coordinates ) and the magnetic field perpendicular to the optical axis (equatorial view). For a polar view such as used in \cite{Matsukevich-08} the collection fraction is given by Eq. \ref{eqCaptureFracSigmaPolarPiEquator} for  $\sigma^{\pm}$ transitions and Eq. \ref{eqCaptureFracPiPolarPi} for $\pi$ transitions. For an equatorial view, such as used in \cite{Blinov-04, Moehring-07}, the collection fraction is given by Eq. \ref{eqCaptureFracSigmaEquator} for $\sigma^{\pm}$ transitions and given by Eq. \ref{eqCaptureFracSigmaPolarPiEquator} for the $\pi$ transition. Even though the emission pattern is different for polar/$\sigma$ and equatorial/$\pi$, the fraction of light captured in an acceptance cone of angle $\theta_{m}$ is identical. The numerical aperture approximations are valid within 2\% for NA$< 0.8$.

\begin{eqnarray}
f_{p\sigma,e\pi}(\theta_m) &=  \frac{1}{2} - \frac{7}{16}\cos{\theta_{m} } - \frac{1}{16}\cos{2\theta_{m}}& \approx \frac{3}{8}\mbox{NA}^2 + \frac{1}{64}\mbox{NA}^6
\label{eqCaptureFracSigmaPolarPiEquator} \\
f_{p\pi}(\theta_m) &=\left( 2 +\cos{\theta_m}\right) \sin^4{\frac{\theta_m}{2}}&\approx \frac{3}{16}\mbox{NA}^4 + \frac{1}{16}\mbox{NA}^6
\label{eqCaptureFracPiPolarPi} \\
f_{e\sigma}(\theta_m) &=\frac{1}{2} - \frac{17}{32}\cos{\theta_{m} } + \frac{1}{32}\cos{2\theta_{m}}&\approx \frac{3}{16}\mbox{NA}^2 + \frac{3}{32}\mbox{NA}^4+\frac{5}{128}\mbox{NA}^6
\label{eqCaptureFracSigmaEquator}
\end{eqnarray}

For capturing photons in the most favorable conditions ( $\sigma^{\pm}$ from a polar perspective, or $\pi$ from an equatorial perspective) the coherent coupling of the characterized PFL is $p_{coh}\ge0.64\%$, given diffraction-limited performance at the effective divergence angle $\theta_e=246$ mrad and the measured 30\% focusing efficiency. A higher-efficiency blazed PFL \cite{Cruz-Cabrera-07} would at least double this to $p_{coh}\ge1.3$\%. The actual $p_{coh}$ may be higher than this estimate as the optimum tradeoff between lower $M^2$ and greater aperture will be determined{\it in situ}. From the measured diffraction efficiency and the lens NA the photon collection efficiency should be $p_{coll}$ =4.6\%. For comparison, in recent remote ion entanglement experiments \cite{Moehring-07, Matsukevich-08} the coherent coupling was $p_{coh}\approx$0.32\%, as estimated from the lens numerical aperture and coupling efficiencies stated in the literature. The robust detection scheme used in these experiments requires interference of two fluorescence photons from the two ions at a beamsplitter and subsequent coincident detection. The photons only interfere if they are in the same spatial mode, making the coherent coupling efficiency $p_{coh}$ rather than the collection efficiency $p_{coll}$ the relevant measure of optical detection effictiveness. For perfect interference, the entanglement rate scales as $p_{coh}^2$, so the use of diffraction-limited high-NA optics such as PFLs promises great benefits.

\subsection{Entanglement with $\sigma^{\pm}$ transitions}
\begin{figure} [htbp]
\centerline{\includegraphics{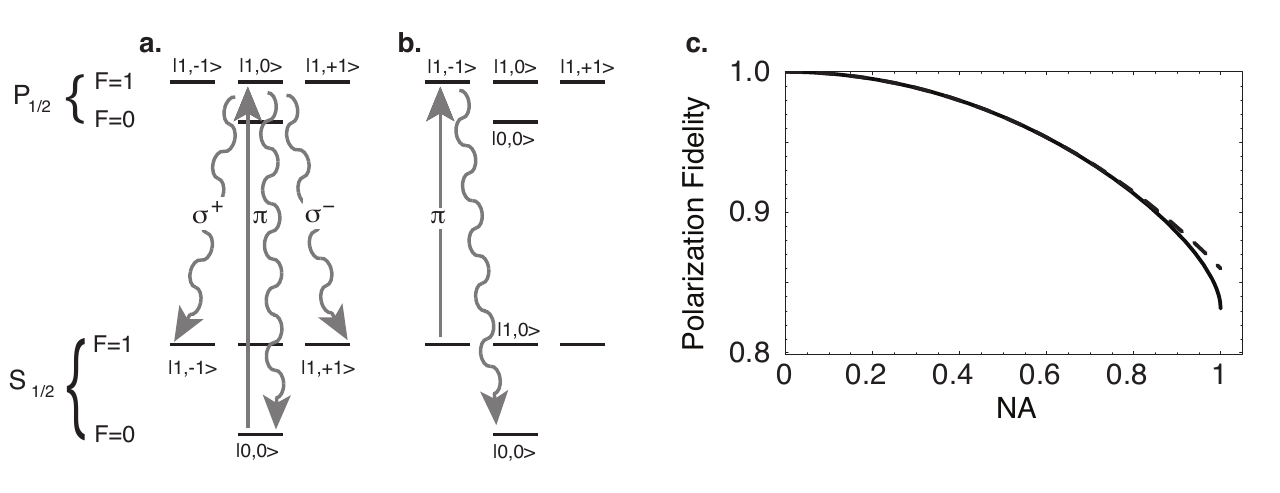}} 
\vspace*{13pt}
\fcaption{a. $\pi$ transition to excited state. Note that selection rules prevent a $\pi$ photon emission at the same frequency as the $\sigma^{\pm}$ photons. b. Optical pumping back to F=0,m$_F$=0 state. For clarity only the primary pumping process for one entangled state is shown.  c. Polarization fidelity of $\sigma^{\pm}$ photons in a polar view as a function of numerical aperture. Minimum fidelity is 0.832 at NA=1.0. Dashed line is the approximation $F(\mbox{NA})\approx1- \mbox{NA}^2/8 -\mbox{NA}^4/96-7\mbox{NA}^6/1536$, valid to $<$1\% for NA$<$0.95.}
\label{FigPolarSigmaEntanglementScheme}
\end{figure}

We now consider a specific ion-photon entanglement scheme based on $\sigma^{\pm}$ Raman transitions in $^{171}$Yb$^+$. This configuration was recently used as part of a demonstration of Bell inequality violation between two remotely entangled ions \cite{Matsukevich-08}. Consider an ion initialized in the S$_{1/2}$ $\left|0, 0\right>$ ($\left|F, m_F\right>$) ground state. A laser drives the $\pi$ polarized transition into the P$_{1/2}$ $\left|1,0\right>$ (Fig. \ref{FigPolarSigmaEntanglementScheme}a) excited state, with three possible decay channels. The ion can return to the $\left|0, 0\right>$ ground state via a Rayleigh scattered $\pi$ polarized photon with 1/3 probability. Alternatively the ion can Raman scatter into the $\left|1, +1\right>$ or $\left|1,-1\right>$ states with a $\sigma^+$ or $\sigma^-$ polarized photon, each with probability 1/3, which is frequency shifted from the excitation laser by 12.6 GHz (the ground state hyperfine splitting in $^{171}$Yb$^+$). A collection optic oriented for viewing parallel to the magnetic field (polar orientation) will gather mostly $\sigma^{\pm}$ photons (Eq. \ref{eqCaptureFracSigmaPolarPiEquator} ) as the dipole radiation pattern of the $\pi$ transition is suppressed in this direction (Eq. \ref{eqCaptureFracPiPolarPi} and Fig \ref{FigLensNACaptureFraction}). In this orientation the $\sigma^{\pm}$ photons appear to be circularly polarized and can be converted to linear polarization with a quarter-wave plate. Further suppression of the unwanted $\pi$ photons at large NA can be obtained with a Fabry-Perot etalon that selectively transmits the Raman shifted $\sigma^{\pm}$ photons. A factor of 1000 in suppression can be obtained with an etalon of finesse 50 and a free spectral range (FSR) of twice the Raman shift (25.2 GHz). After the entangling step an adiabatic microwave sweep can transfer the entangled ion states into the magnetically insensitive (m$_F=$0) clock states for storage or detection. To complete the cycle the ion can be reinitialized to the $\left|0,0\right>$ state with an optical pumping step (Fig. \ref{FigPolarSigmaEntanglementScheme}b).

This scheme generates an entangled ion-photon pair 2/3 of the time for a single scattering cycle and has excellent suppression of unwanted $\pi$ photons. An additional source of error that becomes prominent with large numerical aperture optics is the reduction in polarization contrast (blurring) between $\sigma^+$ and $\sigma^-$ photons at large angles.  The polarization fidelity of a photon emitted at an angle $\theta$ from the optical axis drops as $\sqrt{1-\frac{1}{2}\sin^2\theta}$. The polarization fidelity of captured photons as a function of NA is plotted in Fig. \ref{FigPolarSigmaEntanglementScheme} c. and is determined by weighting the emission angle dependent fidelity by the emission probability distribution for a $\sigma$ transition. Because of this blurring, polarization fidelity greater than 99\% is limited to NA$<0.27$, while a 90\% fidelity requires NA $<0.85$ without additional steps. Fortunately the tradeoff between collection efficiency and entanglement fidelity can be removed using a similar technique as that used for $\pi$ light suppression. The application of a sufficient magnetic field to resolve the Zeeman levels allows for filtering with Fabry-Perot etalons. A reduction of 100 in this error rate can be obtained with an etalon of finesse 16 and 320 MHz FSR with ions in a 67 gauss field with a 160 MHz Zeeman splitting. A drawback to this resolved Zeeman splitting approach is the increased complexity in driving the cooling and readout transitions. In implementing this approach it will likely be easier to choose a magnetic field strength such that a single etalon can remove both $\pi$ light and the unwanted $\sigma$ light. If purely polarization basis photonic qubits are desired for subsequent processing, acousto-optic modulators can be used to remove the Zeeman splitting frequency shift.

\subsection{Chromatic aberration in phase Fresnel lenses}
The large change in focal length with wavelength or chromatic dispersion of phase Fresnel lenses is well known to optical engineers. It cannot be practically compensated with conventional refractive optics for high NA PFLs. While this is not an issue for photonic qubits encoded in the polarization basis \cite{Blinov-04}, it does present a concern for those that have been encoded with the hyperfine frequency difference \cite{Moehring-07}. For a small change in wavelength the focal shift of a PFL is $\Delta f \approx f_0 \Delta \lambda/\lambda_0$.

In $^{171}$Yb$^+$ the 12.6 GHz ground state hyperfine splitting corresponds to a fractional difference in wavelength of $\Delta \lambda/\lambda=15\times10^{-6}$ on the 369.5 nm transition line. For the characterized f=3 mm PFL this results in a small but non-negligible focal shift of 47 nm or  ~5\% of the Rayleigh range. The chromatic focal shift $\Delta f$ is less than the nominal Rayleigh range (or depth of focus) $z_R=4\lambda/\left( \pi \mbox{NA}^2\right)$ for PFLs with focal length $f< 4\lambda^2/\left( \pi\Delta \lambda \mbox{NA}^2 \right)$. For $^{171}$Yb$^+$ ions with an 0.9 NA lens this criteria is satisfied for focal lengths less than 39 mm, and hence should not present a problem for integrated PFL arrays. For large departures from the design wavelength the aberration corrections in the PFL design cease to be valid and a chromatically dependent spherical aberration limits the spot size. We have observed for $\Delta \lambda$= 1 nm that this effect limited the beam quality factor to $M^2\geq1.7$, well above the diffraction limit.

\section{Fresnel lenses for large-scale ion-trap quantum computing}

Most proposals for large-scale ion-trap quantum computing (QC) envisage the use of a large array of ion traps and the collection of ion fluorescence from many sites of the trap array \cite{Kielpinski-02,Wineland-Meekhof-expt-issues-ion-QC,  Duan-Monroe-probabilistic-ion-photon-QC, Duan-fast-ion-gate-QC}. Trapping and transport of ions in microfabricated trap arrays is now routine \cite{Leibfried-07} and current trap fabrication technology is suited to the production of extremely large trap arrays \cite{Seidelin-Wineland-surface-trap}. These same sophisticated ion-trap experiments collect ion fluorescence using complex, bulky multi-element objective lenses that must be aligned manually, a technology that is hardly scalable at all. A new approach will be required for highly parallel fluorescence collection.

Arrays of phase Fresnel lenses are an excellent candidate for scalable fluorescence collection optics. We have shown how single PFLs can significantly improve collection efficiency and coherent mode conversion over present-day optics. In addition, large PFL arrays can be microfabricated on a single substrate, so that only a single alignment step is needed for high-efficiency collimation of light from a large number of trap sites. These ideas were already realised for optical communications as long ago as the 1980s, with the fabrication of high-density optical interconnects based on PFL arrays \cite{Hornak-PFL-array-fab}. This already mature technology, improved for higher numerical aperture and UV wavelengths, can bring massively parallel fluorescence collection to ion-trap QC.

Alternative methods of fluorescence collection from a trap array include the use of a single multi-element objective to image the entire array at once \cite{Kim-05} or the use of an array of conventional microlenses. In the first case, the semiconductor industry has indeed demonstrated lithographic imaging with submicron resolution over large flat field of 25 mm in diameter \cite{Zeiss-08}, but the large, multi-element imaging lenses involved are the result of an extensive design effort, concentrated at a few wavelengths of interest to that industry, that is unlikely to be replicated in quantum computing research. Indeed, these imaging lenses might even be replaced by PFL arrays in semiconductor lithography itself \cite{Gil-Smith-PFL-array-fabrication}.

High NA refractive microlens arrays appear to share desirable features with PFL arrays, but these devices are fabricated from materials which are either not vacuum compatible or strongly absorbing at the UV wavelengths relevant to ion-trap quantum computing. For materials that meet both the UV and vacuum compatibility criterial, microlens arrays have been demonstrated up to NA of 0.3 \cite{Ottevaere-06}, with diffraction-limited performance below NA of 0.2 \cite{Merenda-07}. The photoresists and epoxies used in fabricating high NA microlenses \cite{Ottevaere-06} are designed to be activated by UV light and hence make them poor choices as materials for UV optics. In addition, while microlenses with NA's of up to 0.85 \cite{Miyashita-05} have been demonstrated, these lenses are composed of only a few refracting surfaces and do not exhibit diffraction-limited performance. The advantages of large NA, diffraction-limited performance, vacuum compatibility, and low UV absorption make PFL arrays a superior choice for massively parallel ion trap QC.

\subsection{Highly parallel, efficient detection for quantum error correction}

Every qubit in a large-scale quantum computer will require frequent and repeated error correction, and thus frequent and accurate measurement, to avoid decoherence. For this reason, calculations of fault-tolerance error thresholds for large-scale QC generally assume that error correction can be applied in parallel to all logical qubits \cite{Steane-error-correction-threshold}. A fault-tolerant ion-trap quantum computer is expected to require photon collection efficiency $p_{coll}$ of at least 5\% with parallel detection at $\sim 10^3$ sites \cite{Steane-07}, corresponding to NA $\geq 0.44$ for an ideal lens under the top-hat criterion. Here we have assumed a detection quantum efficiency of 20\%, as is typical for ion-trap experiments \cite{Moehring-07}. \\ \\

PFL arrays are well suited to highly parallel, efficient detection. The e-beam lithography process for PFL array fabrication is extremely flexible in terms of design wavelength and focal length, since these are controlled by changing the groove spacing and depth. The only factors limiting the NA of the PFL array are 1) the distance between detection regions, which sets the maximum useful clear aperture and 2) the minimum allowable distance to the PFL array owing to proximity effects on the ion-trapping fields, which sets the minimum focal length. A trap optimally designed for ion shuttling should have segments of length $\approx d/2$, where $d$ is the distance to the nearest electrode \cite{Home-Steane-transport-electrode-config}. At least seven segments per trap site are required for a square 2D trap array, since each trap site should be able to split an ion crystal and each junction must be able to direct ions along any desired path. An efficient quantum error-correcting code requires less than 1/5 of the physical qubits to be measured \cite{Steane-07}, so detection is required at less than 1/5 of the trap sites. We then find a separation of $> 7.8 d$ between detection regions. As for the proximity effects, the fused silica used for our test PFL is known to have low RF loss, and in the absence of surface charging, the effect of a lossless dielectric surface on the local RF field must be less than that of a conducting surface at the same location. We take a focal length for the Fresnel lens of $\sim 3 d$, which is sufficiently large that proximity effects on the trap fields are relatively minor \cite{Keller-Walther-THESIS}. We then find that the useful NA of a PFL array is at least 0.6 (the NA of our test lens), and can be substantially higher. At NA = 0.6, the diffraction efficiency can easily exceed 60\% \cite{Cruz-Cabrera-07}, so we anticipate a collection efficiency of $p_{coll} \geq 8$\% at each site, sufficient for fault-tolerant QC. In contrast, microlens arrays are limited to a collection efficiency of 3.4\%, assuming a maximum NA= 0.3 and no losses.

\subsection{Coherent mode conversion for ion-photon networking}

We now consider probabilistic quantum networking of ions with photons \cite{Moehring-Monroe-ion-photon-network-rev}, which exploits the entanglement between the ion state and the outgoing photon state created by spontaneous emission. Recent experiments have created remote ion-ion entanglement by the probabilistic detection of individual fluorescence photons emitted by two widely separated ions \cite{Moehring-07,Matsukevich-08}. An array of small ion-trap quantum registers, supplemented by remote entanglement of this kind, could support fault-tolerant QC without the need for ion shuttling \cite{Duan-Monroe-probabilistic-ion-photon-QC}.

The phase Fresnel lens is an outstanding technology for massively parallel ion-photon networking because of its unique potential for high $p_{coh}$. Remote entanglement is a scarce resource in large-scale QC, so we anticipate that a networking architecture will favor high $p_{coh}$ over high density of detection regions, especially since ions need not be shuttled between sites for logic operations. In the coincident-detection scheme described above, the leading sources of error are the imperfect polarization of the excitation laser and the imperfect filtering of photons, and the estimated error rate is less than 0.01 even at NA close to 1. Rigorous coupled-wave calculations for a four-level PFL show that the diffraction efficiency can exceed 50\% at an NA of 0.8 \cite{Cruz-Cabrera-07}, for which the collection efficiency of $\sigma_+/\sigma_-$ photons is 25\%. Our measured $M^2$ at NA = 0.64 is less than 1.1, and the imaging resolution of PFLs at 400 nm conforms to the theoretical prediction for NA up to 0.85 \cite{Menon-06}, so we conservatively estimate $M^2 < 1.5$ at NA of 0.8. Then $p_{coh} = 6$\%, giving a 200-fold increase of entanglement rate over \cite{Matsukevich-08}.

\section{Conclusion}
\noindent
Large numerical aperture optics are crucial to scaling up ion-trap quantum computing. We have shown how phase Fresnel lens arrays are a superior alternative to conventional multi-element lenses or refractive microlens arrays for integration with chip type ion traps. In particular, coincident-detection-based ion-photon entanglement schemes benefit from PFLs since the entanglement rate scales as $\mbox{NA}^4$ and depends on high spatial mode quality. We have demonstrated the optical viability of a PFL as a large NA collection and coherent coupling optic for ion-trap quantum computing with the measurement of a diffraction-limited sub-micron spot. The chromatic aberrations in PFLs were quantified and found not to limit performance at high NA for typical trap configurations. In addition we have proposed two fidelity enhancement techniques to the ion-photon entanglement scheme based on $\sigma^{\pm}$ polarized photons from a Raman transition.

\nonumsection{Acknowledgments}
\noindent

We would like to thank the Australian Research Council (ARC) for their support of this research through Discovery Project grants DP0773354 (Kielpinski) and DP0877936 (Streed) as well as Prof. Howard Wiseman's Federation Fellowship grant FF0458313. This work was also supported by the US Air Force Office of Scientific Research under contract FA4869-06-1-0045. E. Streed acknowledges the support of an Australian Postdoctoral Fellowship from the ARC. We thank Prof. Wolfgang Lange for helpful discussions. Phase Fresnel lenses were fabricated by Margit Ferstl at the Heinrich-Hertz-Institut of the Fraunhofer-Institut f\"{u}r Nachrichtentechnik in Germany.

\nonumsection{References}

\end{document}